\documentclass[a4paper,onecolumn,english,british]{svmult}
\usepackage[T1]{fontenc}
\usepackage[latin9]{inputenc}
\usepackage{bm}
\usepackage{amssymb}
\usepackage{graphicx}

\makeatletter

\usepackage{graphics}

\usepackage{times}

\usepackage{amstext}

\usepackage{bbm}

\makeatother

\usepackage{babel}

\makeatother

\usepackage{babel}
\begin{document}

\title{Gateway schemes of quantum control for spin networks%
\thanks{Cite as: K. Maruyama and D. Burgarth, \textit{Gateway schemes of quantum
control for spin networks}, Chapter 6 in T. Takui, L. J. Berliner,
and G. Hanson (eds.), Electron Spin Resonance (ESR) Based Quantum
Computing, Springer New York, pp 167--192 (2016).%
}}

\author{Koji Maruyama$^{1}$ and Daniel Burgarth$^{2}$}

\maketitle
$^{1}$Department of Chemistry and Materials Science, Osaka City University,
Osaka 558-8585, Japan

\noindent$^{2}$Institute of Mathematics and Physics, Aberystwyth
University, Aberystwyth SY23 3BZ, UK

\section{Motivation and Overview}

Towards the full-fledged quantum computing, what do we need? Obviously,
the first thing we need is a (many-body) quantum system, which is
reasonably isolated from its environment in order to reduce the unwanted
effect of noise, and the second might be a good technique to fully
control it. Although we would also need a well-designed quantum code
for information processing for fault-tolerant computation, from a
physical point of view, the primary requisites are a system and a
full control for it. Designing and fabricating a controllable quantum
system is a hard work in the first place, however, we shall focus
on the subsequent steps that cannot be skipped and are highly nontrivial.

Typically, when attempting to control a many-body quantum system,
every subsystem of it has to be a subject of accurate and individual
access to apply operations and to perform measurements. Such a (near-)
full accessibility leads to a problem of not only technical difficulties,
but also noise (decoherence), as the system can readily interact with
its surrounding environment. In a sense, we are wishing for two inconsistent
demands, namely, being able to manipulate a quantum system fully by
controlling the field parameters while suppressing its interaction
with the field.

A good news is that the technological progress over the last decades
has been so great that we are now able to access and control quantum
systems quite well, provided they are not too large. The coherent
manipulations of small quantum systems, in addition to the observations
of quantum behaviours, have been reported for various systems, e.g.,
NMR/ESR \cite{Vandersypen2005,Vandersypen2001,Laflamme2002,Morton2005},
semiconductor quantum dots \cite{Petta2005,Koppens2006,Hayashi2003},
superconducting quantum bits (\textsl{qubits}) \cite{Nakamura1999,Makhlin2001,Vion2002},
and NV-centres in diamonds \cite{Jelezko2004,Epstein2005}.

Here, we discuss a possible scheme to bridge the gap between what
we wish to achieve and what we can realise today. Namely, we aim at
controlling a given many-body quantum system and identifying it by
accessing only a small subsystem, i.e., \textit{gateway}. Restricting
the size of accessible gateway and minimising the number of control
parameters should be of help in suppressing the effects of noise.

This chapter consists of two parts, each of which is devoted to these
two topics, full quantum control through a gateway and Hamiltonian
identification, respectively. Such situations, in which only a subsystem
is accessible, arise for example in networks of `dark spins' in diamond
and solid state quantum devices\cite{Epstein2005,Neeley2008,Yuasa2009}
as well as spin networks in NMR and ESR setups \cite{Vandersypen2005,Morton2005,Sato2012}.

In the first part, we present how a system can be controlled through
access to a small gateway. Starting with a general argument on the
controllability of a quantum system, we show a possible scheme to
control spin networks under limited access. The two major issues of
our interest in terms of the controllability concern the algebraic
criterion for the form of Hamiltonians and the topological (or graph
theoretical) condition for the choice of gateway. While the consideration
about these aspects will lead to clear insights into the control of
spin-1/2 systems, the theory is general enough to be applied to other
systems we encounter in the lab. We shall also discuss a few issues
related to efficiency, such as, can we compute a pulse sequence for
a certain unitary on the chain by a classical computer within polynomial
time? Or how much time would a unitary require to be performed?

All these discussions on the controllability assume the complete knowledge
of the system Hamiltonian. The second part of this chapter is devoted
to the discussions on how the Hamiltonian can be identified despite
the limited access. Without the knowlege of Hamiltonian, we can never
control a quantum system at will: it will be like going for treasure
hunting without a map and a compass. Having learned the details of
the system Hamiltonian, we then attempt to fully control it, enjoying
the quantumness of the dynamics. Nonetheless, both the full information
acquisition and the full control are still very hard. In addition,
the operational complexity of information acquisition (state and process
tomographies) grows rapidly (exponentially) with respect to the system
size.

Presumably the most straightforward way to estimate the quantum dynamics
is to apply quantum process tomography (QPT), which is a method to
determine a completely positive map $\mathcal{E}$ on quantum states.
The map $\mathcal{E}$ on a state $\rho$ can be written as $\mathcal{E}(\rho)=\sum_{i}E_{i}\rho E_{i}^{\dagger},$
where the operators $E_{i}$ satisfy $\sum_{i}E_{i}^{\dagger}E_{i}=I$
(if $\mathcal{E}$ occurs with unit probability) \cite{Kraus_book1983}.
The complexity of QPT grows exponentially with respect to the system
size; for a $N$ qubit system, we need to specify $2^{4N}$ parameters
for $\mathcal{E}$ and it is an overwhelming task even for small qubit
systems \cite{Schirmer2008,Devitt2006,Young2009}. Moreover, QPT necessitates
estimating all the matrix elements of $\rho$, the state of the whole
system, which is impossible under a restricted access with zero or
little knowledge on the Hamiltonian.

The hardness of the task stems from our complete ignorance about the
nature of the dynamics. However, here we will consider the cases in
which some a priori knowledge or good plausible assumptions are available
to us. In reality, it is natural to have substantial knowledge on
a fabricated physical system, which is the subject of our control,
due to the underlying physics we intend to exploit. Thus, here we
will see how such a priori information on the system can help reduce
the complexity of Hamiltonian identification. We will primarily focus
on the systems consisting of spin-1/2 particles. This is largely because
they have been attracting much attention recently as a promising candidate
for the implementation of quantum computers.

Yet, it would not make much sense if the size of the gateway is comparable
to that of the entire system. From the viewpoint of noise suppression,
the smaller the gateway size, the better. Then how can we find a minimal
gateway that suffices to obtain full knowledge on the system? As we
will see below, the same graph property we introduce in the first
part, i.e., the study of spin network control, comes in to the discussion
as a criterion for estimability of the spin network Hamiltonian.

This Chapter is based on the results from \cite{Burgarth2009b,Burgarth2010,Burgarth2009,Burgarth2009a,Burgarth2011}
as well as some new results.

\part{Indirect control of spin networks}

\section{Reachability in Quantum Control\label{sec:Reachability}}

A central question in control theory is provided a system, typically
described by states, interactions, and our influence on them, to characterize
the operations that can be achieved by suitable controls. In (unitary)
quantum dynamics, the usual setup is a time dependent Hamiltonian
of the form
\begin{equation}
H(t)=H_{0}+\sum_{k}f_{k}(t)H_{k},\label{eq:setup}
\end{equation}
where the time dependence $f_{k}(t)$ can be chosen by the experimentator.
While in usual quantum mechanics we solve the Schrödinger equation
for a given $f_{k}(t)$ to obtain a time evolution unitary $U,$ the
question of control is exactly the inverse: provided a unitary $U,$
is there a control $f_{k}(t)$ which achieves it? The unitaries for
which this is true are called \emph{reachable.}

Given a system (\ref{eq:setup}), how do we characterize the reachable
unitaries? It turns out that it is easier to include those unitaries
which are reachable \emph{arbitrarily well} into our consideration,
and to describe things in terms of \emph{simulable Hamiltonians: }we
call a Hamiltonian $iH$ simulable if $\exp(-iHt)$ is reachable arbitrarily
well for any $t\ge0.$ Clearly, $iH_{0}$ is effectively reachable
by setting $f_{k}\equiv0$ and letting the system evolve for a suitable
time $t.$ We could also set $f_{1}\equiv1$ and all others zero,
and simulate $iH_{0}+iH_{1},$ and so on. Let us call the simulable
set $\mathcal{L}$ and see which rules it obeys:
\begin{enumerate}
\item $A,B\in\mathcal{L}\Rightarrow A+B\in\mathcal{L}:$ this is a simple
consequence of Trotter's formula, which says that by switching quickly
between $A$ and $B$ the system evolves under the average of $A$
and $B.$
\item $A\in\mathcal{L},\alpha>0\Rightarrow\alpha A\in\mathcal{L}:$ this
follows simply from letting a weaker interaction evolve longer to
simulate a stronger one, and vice versa.
\item $A,-A,B,-B\in L\Rightarrow[A,B]\in\mathcal{L}:$ this follows from
a not so well-known variant of Trotter's formula given by
\begin{equation}
\lim_{n\rightarrow\infty}\left(e^{Bt/n}e^{At/n}e^{-Bt/n}e^{-At/n}\right)^{n^{2}}=e^{-[A,B]t^{2}}\label{eq:trotter2}
\end{equation}

\item $A\in\mathcal{L}\Rightarrow-A\in\mathcal{L}:$ This is a property
which heavily relies on finite dimensions, where the quantum recurrence
theorem holds,
\begin{equation}
\forall\epsilon,t>0\exists T>t:\quad||e^{-AT}-1||\le\epsilon\label{eq:recurrence}
\end{equation}
which implies $e^{-A(T-t)}\approx e^{+At}.$
\end{enumerate}
If we combine all the above properties we find that the simulable
set obeys exactly the properties of a Lie algebra over the reals.
This is very useful; in particular, if through rules 1-4 \emph{arbitrary
}Hamiltonians can be simulated, then likewise arbitrary unitaries
are reachable: the system is \emph{fully controllable~}~\cite{Lloyd2004,Albertini2002,Schirmer2001}\emph{
}(in fact, this condition is necessary and sufficient)\emph{. }It
was shown by Lloyd that it is a generic property: in fact two randomly
chosen Hamiltonians are universal for quantum computing almost surely.
We will not prove this here as we are going to show something stronger:
a randomly chosen pair of two-body qubit Hamiltonians is universal
for quantum computing almost surely. That is, Lloyd's result holds
even when restricting ourselves to \emph{physical }Hamiltonians.

\section{Indirect Control\label{sec:IndirectContrl}}

The above equations do not yet take into account the structure of
the controls. As discussed in the introduction, it is interesting
to consider the case of composite system $V=C\bigcup\overline{C}$
where only a part $C$ of the system is controlled, while the remainder
$\overline{C}$ is completely untouched. In the light of Eq.~(\ref{eq:setup})
this means that $H_{k}=h_{C}^{(k)}\otimes1_{\overline{C}}.$ Control
is mediated to $\overline{C}$ only through the \emph{drift }$H_{0}=H_{V},$
which acts on $C$ and $\overline{C}$. If through $H_{V}$ the whole
system is controllable, it means that we have a case of \emph{weak
}controllability: the controls $H_{k}$ do not themselves generate
all Hamiltonians, the drift evolution is necessary. This implies that
$H_{V}$ sets a time limit for how quickly the system can be controlled.
It also reveals many-body properties of $H_{V}$ and is therefore
interesting from a fundamental perspective.

The question is, given $H_{V}$ and a split of the system into $C\overline{C},$
how can we decide if the system is controllable? Is the general result
by Lloyd still correct when restricting ourselves to such a split,
and to a physically realistic $H_{V}?$ In the following, we will
aim to answer both questions.

Using the results from the last section, $V$ is controllable if and
only if
\begin{equation}
\left\langle iH_{V},\mathcal{L}(C)\right\rangle =\mathcal{L}(V),\label{eq:imp02}
\end{equation}
where, for the sake of simplicity, we have assumed the $ih_{C}^{(k)}$'s
to be generators of the local Lie algebra ${\cal L}(C)$ of $C$ and
where we use the symbol $\left\langle \mathcal{A},\mathcal{B}\right\rangle $
to represent the algebraic closure of the operator sets $\mathcal{A}$
and $\mathcal{B}$. ${\cal L}(V)$ denotes the full Lie algebra of
the composite system $V.$ The condition~(\ref{eq:imp02}) can be
tested numerically only for relatively small systems. It becomes impractical
instead when applied to large many-body systems where $V$ is a collection
of quantum sites (e.g. spins) whose Hamiltonian is described as a
summation of two-sites terms. For such configurations, a graph theoretical
approach is more fruitful.

\section{Graph infection\label{sec:Graph-infection}}

The proposed method exploits the topological properties of the graph
defined by the coupling terms entering the many-body Hamiltonian $H_{V}$.
This allows us to translate the controllability problem into a simple
graph property, \textit{infection }\cite{Burgarth2007,Severini2008,Alon}.
In many-body quantum mechanics this property has many interesting
consequences on the controllability and on relaxation properties of
the system~\cite{Burgarth2007,Burgarth2009b}. Also, the same property,
also called \textit{zero-forcing}, has been studied in fields of mathematics,
e.g., graph theory, in a different context \cite{Barioli2010}. Let
us start reviewing this infection property for the most general setup,
which will show more clearly where the topological properties come
from.

\begin{figure}
\includegraphics[scale=0.45]{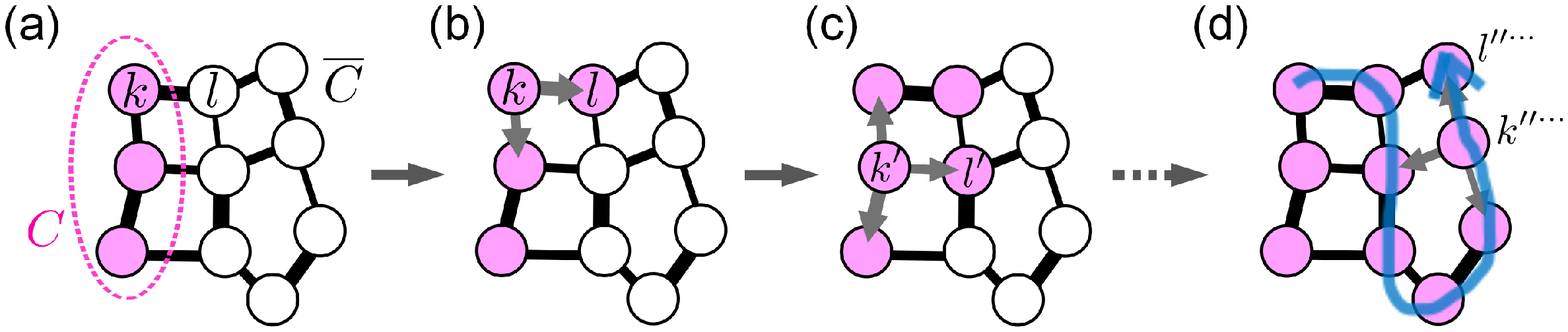}

\caption{\emph{\label{fig:infection}}An example of graph infection. (a) Initially,
three coloured nodes in the region $C$ are `infected'. As the node
$l$ is the only one uninfected node among the neighbours of $k,$
it becomes infected as in (b). (c) Similarly, $l^{\prime}$ becomes
infected by $k^{\prime}$. (d) Eventually all nodes will be infected
one by one.}
\end{figure}

The infection process can be described as follows. Suppose that a
subset $C$ of nodes of the graph is {}``infected'' with some property.
This property then spreads, infecting other nodes, by the following
rule: an infected node infects a {}``healthy'' (uninfected) neighbour
if and only if it is its \emph{unique} healthy neighbour. If eventually
all nodes are infected, the initial set $C$ is called \emph{infecting}.
Figure \ref{fig:infection} would be helpful to grasp the picture.

Note that the choice of $C$ that infects $V$ is not unique. Though
we are interested in small $C,$ finding the smallest one is a nontrivial,
and indeed hard, problem. Nevertheless, from a pragmatic point of
view, the number of nodes we consdier for the purpose of quantum computing
would not be too large to deal with as a graph problem.

\section{Controllability of spin networks}

The link to quantum mechanics is that each node $n$ of the graph
has a quantum degree of freedom associated with the Hilbert space
$\mathcal{H}_{n}$, which describes the $n$-th site of the many-body
system $V$ we wish to control. The coupling Hamiltonian determines
the edges through
\begin{equation}
H_{V}=\sum_{(n,m)\in E}H_{nm}\;,\label{eq:ham}
\end{equation}
where $H_{nm}=H_{mn}$ are some arbitrary Hermitian operators acting
on $\mathcal{H}_{n}\otimes\mathcal{H}_{m}$. Within this context we
call the Hamiltonian~(\ref{eq:ham}) \emph{algebraically propagating}
iff for all $n\in V$ and $(n,m)\in E$ one has,
\begin{equation}
\left\langle \left[iH_{nm},\mathcal{L}(n)\right],\mathcal{L}(n)\right\rangle =\mathcal{L}(n,m),\label{eq:important}
\end{equation}
where for a generic set of nodes $P\subseteq V$, ${\cal L}(P)$ is
the Lie algebra associated with the Hilbert space $\bigotimes_{n\in P}\mathcal{H}_{n}$~%
\footnote{Note that the condition (\ref{eq:important}) is a stronger property
than the condition of controlling $n,m$ by acting on $n$. According
to Eq.~(\ref{eq:imp02}) the latter in fact reads $\left\langle iH_{nm},\mathcal{L}(n)\right\rangle =\mathcal{L}(n,m)$,
which is implied by Eq.~(\ref{eq:important}).%
}. The graph criterion can then be expressed as follows:
\begin{description}
\item [{Theorem:}] \emph{Assume that the Hamiltonian~(\ref{eq:ham}) of
the composed system $V$ is algebraically propagating and that $C\subseteq V$
infects $V$. Then $V$ is controllable acting on its subset $C$.}
\item [{Proof:}] To prove the theorem we have to show that Eq.~(\ref{eq:imp02})
holds, or equivalently that $\mathcal{L}(V)\subseteq\left\langle iH_{V},\mathcal{L}(C)\right\rangle $
(the opposite inclusion being always verified). By infection there
exists an ordered sequence $\{P_{k};k=1,2,\cdots,K\}$ of $K$ subsets
of $V$
\begin{equation}
C=P_{1}\subseteq P_{2}\subseteq\cdots\subseteq P_{k}\subseteq\cdots\subseteq P_{K}=V\;,\label{eq:sequence}
\end{equation}
such that each set is exactly one node larger than the previous one,
\begin{equation}
P_{k+1}\backslash P_{k}=\left\{ m_{k}\right\} ,\label{eq:oneneigh}
\end{equation}
and there exists an $n_{k}\in P_{k}$ such that $m_{k}$ is its unique
neighbor \emph{outside} $P_{k}:$
\begin{equation}
N_{G}(n_{k})\cap V\backslash P_{k}=\left\{ m_{k}\right\} ,\label{eq:unique}
\end{equation}
with $N_{G}(n_{k})\equiv\{n\in V|(n,n_{k})\in E\}$ being the set
of nodes of $V$ which are connected to $n_{k}$ through an element
of $E$. The sequence $P_{k}$ provides a natural structure on the
graph which allows us to treat it almost as a chain. In particular,
it gives us an index $k$ over which we will be able to perform inductive
proofs showing that $\mathcal{L}(P_{k})\subseteq\left\langle iH_{V},\mathcal{L}(C)\right\rangle $.
\end{description}
Basis: by Eq.~(\ref{eq:sequence}) we have $\mathcal{L}(P_{1})=\mathcal{L}(C)\subseteq\left\langle iH_{V},\mathcal{L}(C)\right\rangle $
. Inductive step: assume that for some $k<K$
\begin{equation}
\mathcal{L}(P_{k})\subseteq\left\langle iH_{V},\mathcal{L}(C)\right\rangle .\label{eq:ind}
\end{equation}
We now consider $n_{k}$ from Eq.~(\ref{eq:unique}). We have $\mathcal{L}(n_{k})\subset\mathcal{L}(P_{k})\subseteq\left\langle iH_{V},\mathcal{L}(C)\right\rangle $
and
\begin{eqnarray*}
\left[iH_{n_{k},m_{k}},\mathcal{L}(n_{k})\right]=\left[iH_{V},\mathcal{L}(n_{k})\right]-\sum_{m}\left[iH_{n_{k},m},\mathcal{L}(n_{k})\right],
\end{eqnarray*}
where the sum on the right hand side contains only nodes from $P_{k}$
by Eq.~(\ref{eq:unique}). It is therefore an element of $\mathcal{L}(P_{k})$.
The first term on the right hand side is a commutator of an element
of $\mathcal{L}(P_{k})$ and $iH_{V}$ and thus an element of $\left\langle iH_{V},\mathcal{L}(C)\right\rangle $
by Eq.~(\ref{eq:ind}). Therefore $\left[iH_{n_{k},m_{k}},\mathcal{L}(n_{k})\right]\subseteq\left\langle iH_{V},\mathcal{L}(C)\right\rangle $
and by algebraic propagation Eq.~(\ref{eq:important}) we have
\begin{eqnarray*}
\left\langle \left[iH_{n_{k},m_{k}},\mathcal{L}(n_{k})\right],\mathcal{L}(n_{k})\right\rangle =\mathcal{L}(n_{k},m_{k})\subseteq\left\langle iH_{V},\mathcal{L}(C)\right\rangle .
\end{eqnarray*}
But $\left\langle \mathcal{L}(P_{k}),\mathcal{L}(n_{k},m_{k})\right\rangle =\mathcal{L}(P_{k+1})$
by Eq.~(\ref{eq:oneneigh}) so $\mathcal{L}(P_{k+1})\subseteq\left\langle iH_{V},\mathcal{L}(C)\right\rangle $.
Thus by induction
\begin{equation}
\mathcal{L}(P_{K})=\mathcal{L}(V)\subseteq\left\langle iH_{V},\mathcal{L}(C)\right\rangle \subseteq\mathcal{L}(V).\qquad\blacksquare\newline
\end{equation}

The above theorem has split the question of algebraic control into
two separate aspects. The first part, the algebraic propagation Eq.~(\ref{eq:important})
is a property of the coupling that lives on a small Hilbert space
$\mathcal{H}_{n}\otimes\mathcal{H}_{m}$ and can therefore be checked
easily numerically. The second part is a topological property of the
(classical) graph. An important question arises here if this may be
not only a sufficient but also necessary criterion. As we will see
below, there are systems where $C$ does not infect $V$ but the system
is controllable \emph{for specific coupling strengths}. However the
topological stability with respect to the choice of coupling strengths
is no longer given.

An important example of the above theorem are systems of coupled spin-$1/2$
systems (qubits). We consider the two-body Hamiltonian given by the
following Heisenberg-like coupling,
\begin{equation}
H_{nm}=c_{nm}\left(X_{n}X_{m}+Y_{n}Y_{m}+\Delta Z_{n}Z_{m}\right)\;,\label{equat1}
\end{equation}
where the $c_{nm}$ are arbitrary coupling constants, $\Delta$ is
an anisotropy parameter, and $X$, $Y$, $Z$ are the standard Pauli
matrices. The edges of the graph are those $(n,m)$ for which $c_{nm}\neq0$.

To apply our method we have first shown that the Heisenberg interaction
is algebraically propagating. In this case the Lie algebra ${\cal L}(n)$
is associated to the group $\mbox{su}(2)$ and it is generated by
the operators $\{iX_{n},iY_{n},iZ_{n}\}$. Similarly the algebra ${\cal L}(n,m)$
is associated with $\mbox{su}(4)$ and it is generated by the operators
$\{iX_{n}I_{m},iX_{n}X_{m},iX_{n}Y_{m},\cdots,iZ_{n}Z_{m}\}$. The
identity (\ref{eq:important}) can thus be verified by observing that
\begin{eqnarray*}
\left[X_{n},H_{nm}\right] & = & Z_{n}Y_{m}-Y_{n}Z_{m}\\
\left[Z_{n},Z_{n}Y_{m}-Y_{n}Z_{m}\right] & = & X_{n}Z_{m}\\
\left[Y_{n},X_{n}Z_{m}\right] & = & Z_{n}Z_{m}\\
\left[X_{n},Z_{n}Z_{m}\right] & = & Y_{n}Z_{m},
\end{eqnarray*}
where for the sake of simplicity irrelevant constants have been removed.
Similarly using the cyclicity $X\rightarrow Y\rightarrow Z\rightarrow X$
of the Pauli matrices we get,
\begin{eqnarray*}
X_{n}Z_{m} & \rightarrow & Y_{n}X_{m}\rightarrow Z_{n}Y_{m}\\
Z_{n}Z_{m} & \rightarrow & X_{n}X_{m}\rightarrow Y_{n}Y_{m}\\
Y_{n}Z_{m} & \rightarrow & Z_{n}X_{m}\rightarrow X_{n}Y_{m}.
\end{eqnarray*}
Finally, using
\[
\left[Z_{n}Z_{m},Z_{n}Y_{m}\right]=X_{m}\;,
\]
and cyclicity, we obtain all $15$ basis elements of ${\cal L}(n,m)$
concluding the proof. According to our Theorem we can thus conclude
that \emph{any} network of spins coupled through Heisenberg-like interaction
is controllable when operating on the subset $C$, if the associated
graph can be infected. In particular, this shows that Heisenberg-like
chains with arbitrary coupling strengths admits controllability when
operated at one end (or, borrowing from~\cite{Lloyd2004}, that the
end of such a chain is a universal quantum interface for the whole
system).

\section{General two-body qubit Hamiltonians}

Using the graph criterion we found that the dynamical Lie algebra
for a Heisenberg spin chain with full local control on the first site
\begin{equation}
H_{\mathrm{Hsbg}}+g(t)Y_{1}+f(t)Z_{1}\label{eq:c1}
\end{equation}
is $\mbox{su}(2^{N})$, where $H_{\mathrm{Hsbg}}$ is the Hamiltonian
describing the Heisenberg-type interaction, $H_{\mathrm{Hsbg}}=\sum_{(n,m)\in E}H_{nm}$
with $H_{nm}$ in Eq. (\ref{equat1}). We can also see that the algebra
generated by
\begin{equation}
H_{\mathrm{Hsbg}}+Y_{1}+f(t)Z_{1}\label{eq:c2}
\end{equation}
is $\mbox{su}(2^{N})$.

Extending further, we can consider the Lie algebra generated by $A=H_{\mathrm{Hsbg}}+Y_{1}$
and $B=Z_{1}+1.$ Because $X_{1}=p(A,Z_{1}),$ where $p$ is a (Lie)
polynomial in $A$ and $Z_{1},$ replacing $Z_{1}$ with $Z_{1}+1$
we obtain $p(A,Z_{1}+1)=X_{1}+c1.$ Commuting with $B$ we find that
$Y_{1}$ and therefore also $Z_{1}$ and $1$ seperately are in the
algebra generated by $A$ and $B.$ This has an interesting implication
- namely, that the two Hamiltonians $A=H_{\mathrm{Hsbg}}+Y_{1}$ and
$B=Z_{1}+1$ generate $u(2^{N})$. These are \emph{physical }Hamiltonians,
because they consist of two-body interactions only. The fact that
such pair exists can be used to prove that \emph{almost all pairs
of two-body qubit Hamiltonians }are universal: to do so, we first
observe that we can construct a basis of $u(2^{N})$ through repeated
commutators and linear combinations of $A$ and $B:$
\[
\mbox{u}(2^{N})=\mbox{span}\left\{ p_{1}(A,B),\ldots,p_{2^{2N}}(A,B)\right\}
\]
where the $p_{k}$ are (Lie) polynomials in $A$ and $B.$ The fact
that this is a basis can be expressed equivalently through
\begin{equation}
D\equiv\det\left\{ |p_{1}),\ldots,|p_{2^{2N}})\right\} \neq0,\label{eq:det}
\end{equation}
where $|p_{k})$ is the vector corresponding to the matrix $p_{k}(A,B).$
Now, parametrizing $A$ and $B$ through
\begin{eqnarray}
A & = & \sum_{n,m,\alpha,\beta}a_{\alpha\beta nm}\sigma_{n}^{\alpha}\sigma_{m}^{\beta}\label{eq:exp1}\\
B & = & \sum_{n,m,\alpha,\beta}b_{\alpha\beta nm}\sigma_{n}^{\alpha}\sigma_{m}^{\beta}\label{eq:exp2}
\end{eqnarray}
with $\sigma_{n}^{(0,1,2,3)}\equiv(1_{n},X_{n},Y_{n},Z_{n})$ we can
expand $D$ in Eq.~(\ref{eq:det}) as a multinomial in $a_{\alpha\beta nm}$
and $b_{\alpha\beta nm}$. Our result implies that this multinomial
is not identical to zero, and therefore its roots have measure zero.
Therefore the set of parameters $(a_{\alpha\beta nm},b_{\alpha\beta nm})$
for which the system is not controllable is of measure zero. But the
parametrization (\ref{eq:exp1}) holds for arbitrary two-body qubit
Hamiltonians, which concludes the argument. We note that this argument
is easily extended to general many-body Hamiltonians.

\section{Efficiency considerations}

The above results are interesting from the theoretical point of view;
however, can they be practically useful from the quantum computing
perspective? The two main problems we need to contemplate before attempting
to build a large quantum computer using quantum control are as follows.
First, the precise sequence of actual controls (or `control pulses')
are generally not computable without already simulating the whole
dynamics. We need to find an \emph{efficient }mapping from the quantum
algorithm (usually presented in the gate model) to the control pulse.
Secondly, even if such a mapping can be found, the theory of control
tells us nothing about the overall \emph{duration} of the control
pulses to achieve a given task, and it might take far too long to
be practically relevant.

One approach to circumvent these scaling problems focuses on systems
that are sufficiently small, so that we do not already require a quantum
computer to check their controllability and to design control pulses.
In such a case, the theory of time optimal control \cite{Khaneja2001}
can be used to achieve impressive improvements in terms of total time
or type of pulses required in comparison with the standard gate model.
More complicated desired operations on larger systems are then decomposed
(`compiled') into sequences of smaller ones. Yet, the feasibility
of this approach is ultimately limited by the power of our classical
computers, therefore constrained to low-dimensional many-body systems
only.
\begin{figure}
\includegraphics[scale=0.4]{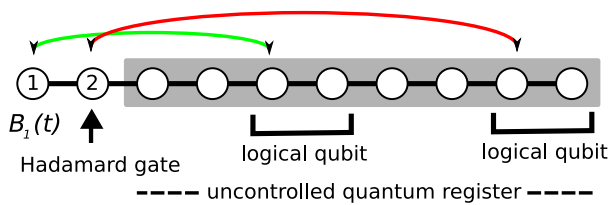}

\caption{\label{fig:Our-scheme-for}(color online) Our approach for universal
quantum computation works on a chain of $N$ spins. By modulating
the magnetic field $B_{1}(t)$ on qubit $1$, we induce information
transfer and swap gates on the chain (red and green lines). The states
of the qubits from the uncontrolled register can be brought to the
controlled part. There, the gates from a quantum algorithm are performed
by local operations. Afterward, the (modified) states are swapped
back into their original position.}
\end{figure}

The goal of this section is to provide an example where one can efficiently
compute control pulses for a large system, using the full Hilbert
space, and to show that the duration of the pulses scales efficiently
(i.e., polynomially) with the system size. We will use a Hamiltonian
that can be efficiently diagonalized for large systems through the
Jordan-Wigner transformation. A similar scheme was developed independently
in~\cite{Kay2010}. The control pulses are applied only to the \emph{first
two} spins of a chain (see Fig.~\ref{fig:Our-scheme-for}). The control
consists of two parts: one where we will use the Jordan-Wigner transformation
to efficiently compute and control the information transfer through
the chain (thus using it as a quantum data bus), and a second part
where we will use some local gates acting on the chain end to implement
two-qubit operations. To be efficiently computable, these local gates
need to be fast with respect to the natural dynamics of the chain.
Combining the two actions allows us to implement any unitary operation
described in the gate model.

More specifically, we consider a chain of $N$ spin-$1/2$ particles
coupled by the Hamiltonian

\[
H=\mbox{\ensuremath{{\textstyle \frac{1}{2}}}}\sum_{n=1}^{N-1}c_{n}[(1+\gamma)XX+(1-\gamma)YY]_{n,n+1}+\sum_{n=1}^{N}B_{n}Z_{n},
\]
where $X,Y,Z$ are the Pauli matrices, the $c_{n}$ are generic coupling
constants, and the $B_{n}$ represent a magnetic field. Variation
of the parameter $\gamma$ encompasses a wide range of Hamiltonians,
including the transverse Ising model ($\gamma=1$; for this case we
require the fields $B_{n}\neq0$) and the $XX$ model ($\gamma=0$).
We assume that the value of $B_{1}$ can be controlled externally.
This control will be used to induce information transfer on the chain
and realize swap gates between arbitrary spins and the two `control'
spins $1,2$ at one chain end. Hence such swap gates are steered \emph{indirectly}
by only acting on the first qubit.

In order to focus on the main idea we now present our method for $\gamma=0$
and $B_{n}=0$ for $n>1$. The general case follows along the same
lines, though more technically involved. Our first task is to show
that by only tuning $B_{1}(t),$ we can perform swap gates between
arbitrary pairs of qubits. First we rewrite the Hamiltonian using
the Jordan-Wigner transformation $a_{n}=\sigma_{n}^{+}\prod_{m<n}Z_{m}$,
into $H=\sum_{n=1}^{N-1}c_{n}\{a_{n}^{\dag}a_{n+1}+a_{n+1}^{\dag}a_{n}\}.$
The operators $a_{n}$ obey the canonical anticommutation relations
$\{a_{n},a_{m}^{\dag}\}=\delta_{nm}$ and $\{a_{n},a_{m}\}=0.$ The
term we control by modulating $B_{1}(t)$ is $h_{1}=Z_{1}=1-2a_{1}^{\dag}a_{1}.$
From Sec \ref{sec:Reachability}, we know that the reachable set of
unitary time-evolution operators on the chain can be obtained from
computing the \emph{dynamical Lie algebra} generated by $ih_{1}$
and $iH.$ It contains all possible commutators of these operators,
of any order, and their real linear combinations. For example, it
contains the anti-Hermitian operators $ih_{12}\equiv[ih_{1},[ih_{1},iH]]/(4c_{1})=i(a_{1}^{\dag}a_{2}+a_{2}^{\dag}a_{1}),$
$ih_{13}\equiv\left[iH,ih_{12}\right]/c_{2}=a_{1}^{\dag}a_{3}-a_{3}^{\dag}a_{1}$
and $ih_{23}\equiv\left[ih_{12},ih_{13}\right]=i(a_{2}^{\dag}a_{3}+a_{3}^{\dag}a_{2}).$
We observe that taking the commutator with $h_{12}$ exchanges the
index $1$ of $h_{13}$ with $2.$ Taking the commutator with $iH$
we find that $ih_{14}\equiv\left[ih_{13},iH\right]+ic_{1}h_{23}-ic_{2}h_{12}=i(a_{1}^{\dag}a_{4}+a_{4}^{\dag}a_{1})$
and $ih_{24}\equiv a_{2}^{\dag}a_{4}-a_{4}^{\dag}a_{2}$ are also
elements of the dynamical Lie algebra. Hence the effect of taking
the commutator with $H$ is raising the index of the $h_{kl}.$ Generalizing
this, we find that the algebra contains the elements $ih_{kl},$ with
$k<l,$ $ih_{kl}\equiv a_{k}^{\dag}a_{l}-a_{l}^{\dag}a_{k}$ for $(k-l)$
even, $ih_{kl}\equiv i(a_{k}^{\dag}a_{l}+a_{l}^{\dag}a_{k})$ for
$(k-l)$ odd, and $h_{k}=Z_{k}=1-2a_{k}^{\dag}a_{k}.$ We thus know
that the time evolution operators $\exp(-\pi ih_{kl}/2)$ (which will
turn out to be very similar to swap gates) can be achieved through
tuning $B_{1}(t).$ The main point is that because both $h_{1}$ and
$H$ are free-Fermion Hamiltonians, the corresponding control functions
can be computed \emph{efficiently} in a $2N$-dimensional space (we
will do so explicitly later). Ultimately, we need to transform the
operators back to the canonical spin representation. Using $a_{k}^{\dag}a_{l}=\sigma_{k}^{-}\sigma_{l}^{+}\prod_{k<j<l}Z_{j},$
we find $\exp(-\pi ih_{kl}/2)=(|00\rangle_{kl}\langle00|+|11\rangle_{kl}\langle11|)\otimes1+(|01\rangle_{kl}\langle10|-|10\rangle_{kl}\langle01|)\otimes L_{kl}$
for $(k-l)$ even. The operator $L_{kl}=\prod_{k<j<l}Z_{j}$ arises
from the non-local tail of the Jordan-Wigner transformation and acts
only on the state of the spins \emph{between} $k$ and $l$, \emph{controlled}
by the state of the qubits $k,j$ in the odd parity sector.

In order to use the chain as a quantum data bus, our goal is to implement
\emph{swap gates} $S_{kl}=|00\rangle_{kl}\langle00|+|11\rangle_{kl}\langle11|+|10\rangle_{kl}\langle01|+|01\rangle_{kl}\langle10|$,
so the fact that we have achieved some modified operators with different
phases on $k,l$ instead, and also the controlled non-local phases
$L_{kl},$ could potentially be worrisome. We will use a method suggested
in \cite{Kay2010} that allows us to tackle these complications. That
is, rather than using the physical qubits, we encode in \emph{logical}
qubits, consisting of two neighbouring physical qubits each. They
are encoded in the odd parity subspace $|01\rangle,|10\rangle$. Although
this encoding sacrifices half of the qubits, the Hilbert space remains
large enough for quantum computation, and the encoding has the further
advantage of avoiding macroscopic superpositions of magnetization,
which would be very unstable. Swapping a logical qubit $n$ to the
control end of the chain then consists of two physical swaps $\exp(-\pi ih_{1\,2n-1}/2)$
and $\exp(-\pi ih_{2\,2n}/2)$. Since both physical swaps give the
same phases, the resulting operation is indeed a full logical swap.
Any single-qubit operation on the logical qubits can be implemented
by bringing the target qubit to the control end, performing the gate
there, and bringing it back again. We could equally decide to perform
single logical qubit gates directly, without bringing them to the
control end. This is possible because $\exp(-ih_{2n-1\,2n}t)$ in
the physical picture translates to $\exp{(-iX_{L,n}t)}$ in the logical
picture, and because $Z_{2n-1}$ is in the algebra generated by $Z_{1}$,
which allows us to perform the operation $\exp{(-iZ_{2n-1}t)}=\exp{(-iZ_{L,n}t)}$.

For quantum computation, we need to be able to perform at least one
entangling two-qubit operation. We choose a controlled-Z operation,
which can be performed by operating only on one physical qubit from
each of the two logical qubits involved; to perform a controlled-Z
between logical qubit $n$ and $m$, we bring the physical qubits
$(2n-1)$ and $(2m-1)$ to the control end, perform a controlled-Z
between them, and bring them back. It is easy to check that again
all unwanted phases cancel out. The controlled-Z could not be efficiently
computed in the interplay with the many-body Hamiltonian $H$, because
it cannot be generated by a quadratic Hamiltonian in the Jordan-Wigner
picture. Therefore, this gate must be implemented on a time-scale
$t_{g}$ much faster than the natural evolution of the chain, i.e.,
$t_{g}\ll\min_{j}\{1/c_{j}\}.$ We can soften this requirement by
using control theory to generate $\exp{(-iZ_{1}X_{2}t)}$ by modulating
$\beta_{1}(t)Y_{1}$ (this is a linear term in the Jordan-Wigner picture),
and then using a fast Hadamard gate on the second site to obtain $\exp{(-iZ_{1}Z_{2}t)}$,
which, together with $\exp{(-iZ_{1}t)}$ and $\exp{(-iZ_{2}t)}$,
gives the controlled-Z gate. This leads to a remarkable conclusion:
besides a fast Hadamard gate on the second qubit, all other controls
required for quantum computation can be computed efficiently within
the framework of optimal control.

The crucial question left open above, is how long does it actually
take to implement the gates? In order to evaluate the efficiency,
we have numerically simulated a range of chain lengths and studied
the scaling of the logical swap operation time $T$ with the (physical)
chain length $N$. We set the coupling strength constant, namely $c_{n}=J\ \forall\ n$.
To provide evidence of a polynomial scaling, we set the simulation
time $T_{N}=N{}^{2},$ (all times are in units of $1/J$ and $\hbar=1$)
and verify for each $N$ that we can find a specific $B_{1}^{\ast}(t)$
that performs the logical swap.

We quantify our success by calculating the error of the operation
$\varepsilon=1-F$, where $F=(|\mathrm{tr}{U^{\dagger}U_{g}}|/N)^{2}$
is the gate fidelity between the time evolution $U$ and the goal
unitary $U_{g}$. This standard choice of fidelity is used for evaluating
generic unitaries, and for our case it is well suited confirming that
the swap gate $S_{kl}\otimes1_{\mbox{rest }}$ acts as the identiy
almost everywhere. However the normalization factor $1/N^{2}$ could
in principle wash out errors in the part of the gate that acts on
qubits $k$ and $l$ only, resulting in the wrong scaling. Therefore,
we checked the reduced gate fidelity (tracing out the rest of the
system) on those qubits alone, finding that its fidelity remains above
$1-10^{-4}$ for all $N$ considered.

The function $B_{1}(t)$ is obtained using techniques from optimal
control theory \cite{Khaneja2001,D'Alessandro2008}. Briefly, the
procedure is as follows: (1) an initial guess is made for the function
$B_{1}(t)$; (2) we run the optimal control algorithm to generate
a new $B_{1}(t)$ which decreases the error of our operation; (3)
steps 1 and 2 are iterated until the final error reaches a preselected
threshold $\varepsilon$. In practice, it suffices to choose a threshold
which is of the same order of magnitude as the error introduced by
the Hadamard gate.

If the algorithm converges for each $N$ and the corresponding $T_{N}$,
giving the optimal pulse sequence $B_{1}^{\ast}(t)$, then we can
assert that the scaling of the operation time is at least as good
as $T_{N}=N{}^{2}$, up to a given precision. Simulating chain lengths
up to $N=40,$ we find that $T_{N}=N^{2}$ can be achieved. We stress
here that the chosen scaling law $T_{N}$ may not necessarily describe
the shortest time on which the physical swap gate can be performed.
However, the dynamical Lie algebra of quasi-free fermions has a dimension
of the order $N^{2}$, indicating that such scaling might be optimal.

\section{Conclusion}

We have seen that control theory provides a powerful framework for
indirect control, and therefore for potential control schemes of large
many-body systems. We could furthermore show that almost all physical
relevant Hamiltonians provide full control, and that at least in some
cases efficient mappings from the gate model to quantum control are
possible. Under which conditions this is true, and if - and how -
such schemes can furthermore be made fault-tolerant in the presence
of noise remains an active area of research. One thing that is clear,
however, is that in order to apply such schemes, good knowledge about
the system Hamiltonian $H_{0}$ is required. In the next part, we
will consider how such knowledge can be obtained using similar indirect
schemes.

\part{Indirect Hamiltonian tomography of spin networks}

\section{The gateway scheme of Hamiltonian tomography\label{sec:The-gateway-scheme}}

It has recently been studied how a priori knowledge on the system
could reduce the complexity of quantum process tomography. A noteworthy
example is the method developed on the basis of compressed sensing
\cite{Gross2010,Shabani2011}, which is originally a scheme to make
a best estimation for all elements of a sparce matrix despite limited
amount of data. Assuming the sparcity under physically plausible settings
has been also a key in other works on indirect Hamiltonian identification.
The results on which we base the most of the following description
exploited the polynomial dimensionality of a subspace we probe \cite{Burgarth2009,Burgarth2009a}.
That is, there is already an exponential reduction for the number
of parameters to be determined. While this assumption puts a condition
on the type of Hamiltonians, it was shown that a larger class of Hamiltonians
(for 1D spin chains) could also be estimated through a gateway Di
Carlo et al. \cite{Franco2009}. We shall see below that this is a
special case of the generic estimation of quadratic Hamiltonians,
which might describe the dynamics of either bosons or fermions on
not only 1D chains but also more general networks.

\begin{figure}
\includegraphics[scale=0.5]{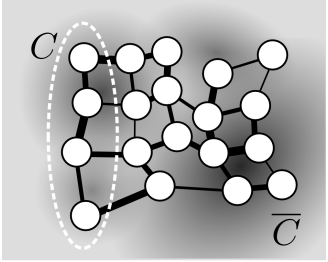}

\caption{\emph{\label{fig:2d}All} coupling strengths (solid lines) and local
magnetic fields (background) of a 2-dimensional network $G=(V,E)$
of spins (white circles) can be estimated \emph{indirectly} by quantum
state tomography on a gateway $C$ (enclosed by the dashed red line).
The coupling strengths and field intensities are represented by the
width of lines and the density of the background colour, respectively.}
\end{figure}

Suppose that we have a network consisting of $N$ spin-1/2 particles,
such as the one in Fig \ref{fig:2d}. Our aim is to estimate all the
non-zero coupling strengths between spins and the intensities of the
local magnetic fields. The assumptions we make are as follows:
\begin{enumerate}
\item The topology of the network is known. That is, information on the
graph $G=(V,E)$ corresponding to the network is available, where
nodes $V$ of the graph correspond to spins and edges $E$ connect
spins that are interacting with each other.
\item The type of the interaction between spins, such as the Heisenberg,
XX, etc., is a priori known.
\item The inhomogeneous magnetic field is applied in the $z$-direction.
\item The values of coupling strengths are all real and their signs are
known.
\end{enumerate}
Assumptions 1 and 2 are the key for reducing the complexity of the
problem. In many experimental situaions, these information are available
due to the conditions for fabrication, albeit a number of exceptions.
In the following, we describe the estimation scheme assuming Hamiltonians
that have the following form:
\begin{eqnarray}
H & = & \sum_{(m,n)\in E}c_{mn}\left(X_{m}X_{n}+Y_{m}Y_{n}+\Delta Z_{m}Z_{n}\right)+\sum_{n\in V}b_{n}Z_{n},\label{eq:simple_ham}
\end{eqnarray}
for simplicity. Here, $X_{m},Y_{m},$ and $Z_{m}$ are the standard
Pauli operators for spin-1/2, $c_{mn}$ are the coupling strengths
between the $m$-th and $n$-th spins, $b_{n}$ are the intensity
of local magnetic field at the site of $n$-th spin, and $\Delta$
is an anisotropy factor that is common for all interacting pairs.

The Hamiltonians of the type of Eq. (\ref{eq:simple_ham}) have a
nice property, $[H,\sum_{n}Z_{n}]=0,$ i.e., the total magnetisation
is preserved under the dynamics generated by $H.$ Thus the whole
$2^{N}$-dimensional Hilbert space is decomposed into the direct sum
of supspaces, each of which corresponds to a specific number of total
magnetisation. For the purpose of Hamiltonian tomography, analysing
the dynamics in the single excitaion sector $\mathcal{H}_{1},$ which
has only a single up spin $|\uparrow\rangle$ among $N$ spins, turns
out be sufficient. We will write a single excitation state as $|\mathbf{n}\rangle\in\mathcal{H}_{1}$
when only the spin $n\in V$ is in $|\uparrow\rangle$ with all others
in $|\downarrow\rangle$ , and $|\mathbf{0}\rangle=|\downarrow...\downarrow\rangle.$
In Sec. \ref{sec:quadratic}, we will treat more general cases, i.e.,
Hamiltonians that do not conserve the total magnetisation, such as
the generic XX- or Ising-type Hamiltonians.

The task of Hamiltonian tomography is to estimate $c_{mn}$ and $b_{n}$
under the limited access to a small gateway $C\subset V$ only. Naturally,
the challenge here is to obtain information about the inaccessible
spins in $\bar{C}\equiv V\setminus C$, which could be a large majority
of $V$. The question is, however, how small can $C$ be such that
we can (in principle) still learn all the couplings and fields in
$V$?

This can be answered by using the \emph{infecting} property, which
has been introduced in Sec \ref{sec:Graph-infection} for a given
graph $G$ and a subset $C\subset V$ of nodes\emph{.} The main theorem
about hamiltonian identification under a limited access can be presented
in terms of the infection property as follows. That is, \emph{if $C$
infects $V,$ then all $c_{mn}$ and $b_{n}$ can be obtained by acting
on $C$ only.} Therefore, $C$ can be interpreted as an upper bound
on the smallest number of spins we need to access for the purpose
of Hamiltonian tomography, i.e., given by the cardinality $|C|$ of
the smallest set $C$ that infects $V.$ To prove this statement,
let us assume that $C$ infects $V$ and that all eigenvalues $E_{j}$
$(j=1,\ldots,|V|$) in \emph{$\mathcal{H}_{1}$} are known. Furthermore,
assume that for all orthonormal eigenstates $|E_{j}\rangle$ in \emph{$\mathcal{H}_{1}$}
the coefficients $\langle\mathbf{n}|E_{j}\rangle$ are known for all
$n\in C.$ We show how these information lead to the full Hamiltonian
identification, and then in Section \ref{sec:tomography} show how
these necessary data, $E_{j}(\forall j)$ in $\mathcal{H}_{1}$ and
$\langle\mathbf{n}|E_{j}\rangle$ for all $j\in{1,...,|V|}$ and all
$n\in C$, can be obtained by simple state tomography experiments.

Observe that the coupling strengths between spins \emph{within} $C$
are easily obtained because of the relation $c_{mn}=\langle\mathbf{m}|H|\mathbf{n}\rangle=\sum E_{k}\langle\mathbf{m}|E_{k}\rangle\langle E_{k}|\mathbf{n}\rangle,$
where we defined $c_{mm}\equiv\langle\mathbf{m}|H|\mathbf{m}\rangle$
for the diagonal terms. Since $C$ infects $V$ there is a $k\in C$
and a $l\in\overline{C}\equiv V\backslash C$ such that $l$ is the
only neighbour of $k$ outside of $C,$ i.e.
\begin{equation}
\langle\mathbf{n}|H|\bm{k}\rangle=0\;\;\forall n\in\overline{C}\backslash\{l\}.\label{eq:null}
\end{equation}
 For an example see Fig. \ref{fig:infection}. Using the eigenequation,
we obtain for all $j$
\[
E_{j}|E_{j}\rangle=H|E_{j}\rangle=\sum_{m\in C}\langle\mathbf{m}|E_{j}\rangle H|\mathbf{m}\rangle+\sum_{n\in V\backslash C}\langle\mathbf{n}|E_{j}\rangle H|\mathbf{n}\rangle.
\]
 Multiplying with $\langle\bm{k}|$ and using Eq.~(\ref{eq:null})
we obtain
\begin{equation}
E_{j}\langle\bm{k}|E_{j}\rangle-\sum_{m\in C}c_{km}\langle\mathbf{m}|E_{j}\rangle=c_{kl}\langle\bm{l}|E_{j}\rangle.\label{eq:mu_H_Ej}
\end{equation}
 By assumption, the left-hand side (LHS) is known for all $j.$ This
means that up to an unknown constant $c_{kl}$ the expansion of $|\bm{l}\rangle$
in the basis $|E_{j}\rangle$ is known. Through normalisation of $|\bm{l}\rangle$
we then obtain $c_{kl}^{2}$, thus $c_{kl}$ (by using the assumed
knowledge on its sign) and hence $\langle\bm{l}|E_{j}\rangle$. Redefining
$C\Rightarrow C\cup\{k\}$, it follows by induction that all $c_{mn}$
are known. Finally, we have
\begin{equation}
c_{mm}=\langle\mathbf{m}|H|\mathbf{m}\rangle=E_{0}-\Delta\sum_{n\in N(m)}c_{mn}+2b_{m},\label{eq:magfield}
\end{equation}
 where $N(m)$ stands for the (directly connected) neighbourhood of
$m,$ and
\begin{equation}
E_{0}=\frac{1}{2}\Delta\sum_{(m,n)\in V}c_{mn}-\sum_{n\in V}b_{n}\label{eq:e0}
\end{equation}
 is the energy of the ground state $|\mathbf{0}\rangle$. Summing
Eq. (\ref{eq:magfield}) over all $m\in V$ and using Eq. (\ref{eq:e0}),
we can have the value of $\sum_{n\in V}b_{n}$, thus that of $E_{0}$
as well, since all other parameters are already known. Then we obtain
the strength of each local magnetic field, $b_{m}$, from Eq. (\ref{eq:magfield}).

\begin{figure}
\includegraphics[scale=0.85]{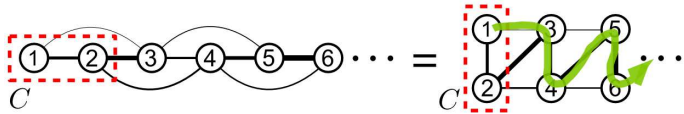}

\caption{\emph{\label{fig:nnn}}An example of graphs for non-nearest neighbour
interactions. The graph for next-nearest interaction (left) can be
infected by $C$ as it is easily seen after deforming (right).}
\end{figure}

An interesting application of the above scheme is a one-dimensional(1D)
spin chain with non-nearest neighbour interactions \cite{Cappellaro2007}.
If spins interact with the next-nearest neighbours in addition to
the nearest ones, the whole graph can be infected by setting the two
end spins as $C$, as shown in Fig.\ref{fig:nnn}. Similarly, if spins
interact with up to $r$-th nearest neighbours, all coupling strengths
can be estimated by including the $r$ spins at the chain end, from
the first to the $r$-th, in $C.$

\section{Data acquisition\label{sec:tomography}}

In order to perform the above estimation procedure, we need to know
the energy eigenvalues $E_{j}$ in $\mathcal{H}_{1}$ and the coefficients
$\langle\mathbf{n}|E_{j}\rangle$ for all $n\in C$ by controlling/measuring
the spins in $C$. Suppose the spin 1 is in $C.$ To start, we initialise
the system as $|\mathbf{0}\rangle$ and apply a fast $\pi/2$-pulse
on the spin 1 to make $\frac{1}{\sqrt{2}}(|\mathbf{0}\rangle+|\mathbf{1}\rangle).$
This can be done efficiently by acting on the spin 1 only; the basic
idea is that by measuring the spin 1, and flipping it quickly every
time when it was found in $|\uparrow\rangle$, the state of the network
becomes $|\mathbf{0}\rangle$ within a polynomial time with respect
to the network size $N=|V|.$ The reason for this is two-fold: the
excitation-preserving property of the Hamiltonian guarantees that
an up-spin cannot be observed more than $N$ times and the propagation
time of up-spins in the network is polynomial in $N$ \cite{Burgarth2007c}.
Then, we perform quantum state tomography on the spin $n\in C$ after
a time lapse $t$. By repeating the preparation and measurements on
spin $n$, we obtain the following matrix elements of the time evolution
operator as a function of $t:$
\begin{equation}
e^{iE_{0}t}\langle\mathbf{n}|U(t)|\mathbf{1}\rangle=\sum_{j}\langle\mathbf{n}|E_{j}\rangle\langle E_{j}|\mathbf{1}\rangle e^{-i(E_{j}-E_{0})t}.\label{eq:fourier}
\end{equation}
 If we take $n=1$ and Fourier-transform Eq.~(\ref{eq:fourier})
we can get information on the energy spectrum in $\mathcal{H}_{1}$.
Up to an unknown constant $E_{0}$, which turns out to be irrelevant,
we learn the values of all $E_{j}$ from the peak positions. The height
of the $j$-th peak gives us the value of $|\langle\mathbf{1}|E_{j}\rangle|^{2}$
for \emph{all} eigenstates. Thanks to the arbitrariness of the global
phase, we can set $\langle\mathbf{1}|E_{j}\rangle>0.$ Hence observing
the \emph{decay/revival} of an excitation at $n=1$ we can learn some
$E_{j}$ and all the $\langle\mathbf{1}|E_{j}\rangle$.

In order to determine $\langle\mathbf{n}|E_{j}\rangle$ for other
$n\in C,$ we prepare a state at 1 and measure at $n.$ Namely, setting
$n(\neq1)$ in Eq. (\ref{eq:fourier}) allows us to extract the coefficient
$\langle\mathbf{n}|E_{j}\rangle$ correctly, including their relative
phase with respect to $\langle\mathbf{1}|E_{j}\rangle$. Continuing
this analysis over all sites in $C,$ we get all information necessary
for the Hamiltonian tomography. It could be problematic if there were
eigenstates in $\mathcal{H}_{1}$ that have no overlap with \emph{any
$n\in C$}, i.e., $\langle\mathbf{n}|E_{j}\rangle=0.$ Fortunately,
such eigenstates do not exist, as shown in~\cite{Burgarth2007}.
Therefore we can conclude that all eigenvalues in the $\mathcal{H}_{1}$
can be obtained. Although tomography cannot determine the extra phase
shift $E_{0},$ it does not affect the estimation procedure (it is
straightforward to check that it cancels out in the above estimation).

Note that in order for the information about $\langle\mathbf{n}|E_{j}\rangle$
$(n\in C)$ to be attained there should be no degeneracies in the
spectrum of Eq. (\ref{eq:fourier}). For example, suppose there are
two orthogonal states $|E_{k}^{(1)}\rangle$ and $|E_{k}^{(2)}\rangle$,
both of which are the eigenstates of $H$ corresponding to the same
eigenvalue $E_{k}$. The height of the peak at $E_{k}$ in the Fourier
transform of $\langle\mathbf{1}|U(t)|\mathbf{1}\rangle$ would be
$|\langle\mathbf{1}|E_{k}^{(1)}\rangle|^{2}+|\langle\mathbf{1}|E_{k}^{(2)}\rangle|^{2}$.
There is no means to esitmate the value of each term from this sum,
let alone the values of $\langle\mathbf{n}|E_{k}^{(1)}\rangle$ and
$\langle\mathbf{n}|E_{k}^{(2)}\rangle$. Also even if there are no
degeneracies, thus if $E_{j}$ are all distinct, the peaks need to
be sharp enought to be resolved. The issues on degeneracies and resolving
peaks are discussed in the following sections \ref{sec:Degeneracy}
and \ref{sec:Degeneracy}.

\section{Degeneracy\label{sec:Degeneracy}}

What if there were degenerate energy levels in the single excitation
subspace $\mathcal{H}_{1}$? While 1D spin chains have no degeneracies
\cite{Gladwell2004}, there could be in general spin networks. Of
course ``exact degeneracy'' is highly unlikely; however approximate
degeneracy could make the scheme less efficient. In this section,
we show that there \emph{always} exists an operator $B_{C}$, which
represents extra fields applied on $C,$ such that it lifts all degeneracies
of $H$ in $\mathcal{H}_{1}$. Because $C$ is only a small subset,
the existence of such an operator is not a trivial problem at all.
In the following, we demonstrate the existence of such a $B_{C}$
by explicitly constructing it, assuming the full knowledge about $H$.
Without the full knowledge of $H$ (as is the case in the estimation
scenario), we could only guess a $B_{C}$ and have it right probabilistically.
Nevertheless, as it is clear from the discussion below, the parameter
space for $B_{C}$ that does not lift all the degeneracies has only
a finite volume. Thus even choosing $B_{C}$ randomly can make the
probability of lifiting the degeneracies to converge exponentially
fast to one.

Once all degeneracies are lifted, we can estimate the full Hamiltonian
$H+\lambda B_{C}\otimes I_{\bar{C}}$ and subtracting the known part
$\lambda B_{C}\otimes I_{\bar{C}}$ completes our identification task.
Here, $\lambda$ is a parameter for the strength of the fields. Although
the extra fields on $C$ do not necessarily have to be a small perturbation,
let us consider a small $\lambda$ to see the effect of $\lambda B_{C}$
on the energy levels, making use of the pertubation theory.

Let us denote the eigenvalues of $H$ as $E_{k}$ and the eigenstates
as $|E_{k}^{d}\rangle,$ where $d=1,\ldots,D(k)$ is a label for the
$D(k)$-fold degenerate states. Let us first look at one specific
eigenspace $\left\{ |E_{k}^{d}\rangle,d=1,\ldots,D(k)\right\} $ corresponding
to an eigenvalue $E_{k}.$ Since the eigenstates considered here are
in $\mathcal{H}_{1}$, we can always decompose them as
\[
|E_{k}^{d}\rangle_{C\bar{C}}=|\bm{\phi}_{k}^{d}\rangle_{C}\otimes|\bm{0}\rangle_{\bar{C}}+|\bm{0}\rangle_{C}\otimes|\bm{\psi}_{k}^{d}\rangle_{\bar{C}},
\]
 where the unnormalised states $|\bm{\phi}_{k}^{d}\rangle_{C}$ and
$|\bm{\psi}_{k}^{d}\rangle_{\bar{C}}$ are in the single excitation
subspace on $C$ and $\overline{C},$ respectively. The state $|\bm{\phi}_{k}^{d}\rangle_{C}\,(\forall d)$
cannot be null, i.e., $|\bm{\phi}_{k}^{d}\rangle_{C}\neq0$, because
if there was an eigenstate in the form of $|\bm{0}\rangle_{C}\otimes|\bm{\psi}_{k}^{d}\rangle_{\bar{C}}$
then applying $H$ repeatedly on it will necessarily introduce an
excitation to the region $C,$ in contradiction to being an eigenstate
\cite{Burgarth2007}. Furthermore, the set $\left\{ |\bm{\phi}_{k}^{d}\rangle_{C},\, d=1,\ldots,D(k)\right\} $
must be linearly independent: for, if there were complex numbers $\alpha_{kd}$
such that $\sum_{d}\alpha_{kd}|\bm{\phi}_{k}^{d}\rangle_{C}=0,$ then
a state in this eigenspace $\sum_{d}\alpha_{kd}|E_{k}^{d}\rangle_{C\bar{C}}=\sum_{d}\alpha_{kd}|\bm{0}\rangle_{C}\otimes|\bm{\psi}_{k}^{d}\rangle_{\bar{C}}$
would be an eigenstate with no excitation in $C,$ again contradicting
the above statement. This leads to an interesting observation that
the degeneracy of each eigenspace can be maximally $|C|-$fold, because
there can be only $|C|$ linearly independent vectors at most in $\mathcal{H}_{1}$
on $C.$ Thus, the minimal infecting set of a graph, a topological
property, is related to some bounds on possible degeneracies, a somewhat
algebraic property of the Hamiltonian.

Now suppose that $\lambda_{k}B_{kC}$ is a perturbation that we will
construct so that it lifts all the degeneracies for an energy eigenvalue
$E_{k}$. Assuming $B_{kC}|\mathbf{0}\rangle_{C}=0$ turns out to
be sufficient for our purpose. The energy shifts due to $B_{kC}$
in the first order are given as the eigenvalues of the perturbation
matrix $_{C\bar{C}}\langle E_{k}^{d}|B_{kC}\otimes I_{\bar{C}}|E_{k}^{d'}\rangle_{C\bar{C}}=_{C}\langle\bm{\phi}_{k}^{d}|B_{kC}|\bm{\phi}_{k}^{d'}\rangle_{C}.$
We want the shifts to be different from each other to lift the degeneracy.
To this end, recall that $\left\{ |\bm{\phi}_{k}^{d}\rangle_{\bar{C}}\right\} $
are linearly independent, which means that there is a similarity transform
$S_{k}$ (not necessarily unitary, but invertible) such that the vectors
$|\bm{\chi}_{k}^{d}\rangle_{C}\equiv S_{k}^{-1}|\bm{\phi}_{k}^{d}\rangle_{C}$
are orthonormal. The perturbation matrix can then be written as $_{C}\langle\bm{\chi}_{k}^{d}|S_{k}^{\dagger}B_{kC}S_{k}|\bm{\chi}_{k}^{d'}\rangle_{C}.$
If we set $S_{k}^{\dagger}B_{kC}S_{k}=\sum_{d}\epsilon_{kd}|\bm{\chi}_{k}^{d}\rangle_{C}\langle\bm{\chi}_{k}^{d}|$
the Hermitian operator
\begin{equation}
B_{kC}\equiv\sum_{d}\epsilon_{kd}\left(S_{k}^{\dagger}\right)^{-1}|\bm{\chi}_{k}^{d}\rangle_{C}\langle\bm{\chi}_{k}^{d}|S_{k}^{-1}\label{eq:BkC}
\end{equation}
 gives us energy shifts $\epsilon_{kd}.$ Therefore, as long as we
choose mutually different $\epsilon_{kd}$ , the degeneracy in this
eigenspace is lifted by $B_{kC}.$ This happens for an arbitrarily
small perturbation $\lambda_{k}.$ So we choose $\lambda_{k}$ such
that the lifting is large while \emph{no new degeneracies} are created,
i.e. $||\lambda_{k}B_{kC}||\neq\Delta E_{ij},$ where $\Delta E_{ij}=E_{i}-E_{j}$
are the energy gaps of $H.$

There may be some remaining degerate eigenspaces of the perturbed
Hamiltonian $H'=H+\lambda_{k}B_{kC}$. Fortunately, since $B_{kC}$
conserves the number of excitations (see Eq. (\ref{eq:BkC})), we
can still consider only $\mathcal{H}_{1}$ and repeat the above procedure
to find operators $B_{k'C}$ to lift degeneracy in each eigenspace
spanned by ${|E_{k^{\prime}}^{d^{\prime}}\rangle}$. Eventually we
can form a total perturbation $B_{C}=\sum_{k}B_{kC}$ that lifts \emph{all}
degeneracies in $\mathcal{H}_{1}$. By perturbation theory a ball
of finite volume around $B_{C}$ has the same property. In practice,
we expect that almost all operators will lift the degeneracy, with
a good candidate being a simple homogeneous magnetic field on $C.$
This is confirmed by numerical simulations \cite{Burgarth2009a}.

\section{Efficiency\label{sec:efficiency}}

The efficiency of the coupling estimation can be studied using standard
properties of the Fourier transform (see \cite{Bracewell1999} for
an introduction). In experiments, the function $\langle\mathbf{n}|U(t)|\mathbf{m}\rangle(m,n\in C)$
is sampled for descrete times $t_{k}$, rather than for continuous
time $t$, with an interval $\Delta t=t_{k+1}-t_{k}$. Therefore an
important cost parameter is the total number of measured points, being
proportional to the sampling frequency, $f=1/\Delta t$. The minimal
sampling frequency is given by the celebrated Nyquist-Shannon sampling
theorem as $2f_{min}=E_{max}$, where $E_{max}$ is the maximal eigenvalue
of $H$ in the first excitation sector.

Due to decoherence and dissipation, the other important parameter
is the total time length $T(=\mathrm{max}(t_{k}))$ over which the
functions need to be sampled to obtain a good resolution. This is
given by the classical uncertainty principle that states that the
frequency resolution is proportional to $1/T$. Hence the minimal
time duration over which we should sample scales as $T_{min}=1/(\Delta E)_{min}$,
where $(\Delta E)_{min}$ is the minimal gap between the eigenvalues
of the Hamiltonian. Also, in order for all peaks in the Fourier transform
to be resolved, the height of the peaks, which are given by $|\langle\mathbf{n}|E_{j}\rangle\langle E_{j}|\mathbf{m}\rangle|$,
should be high enough. That is, all energy eigenstates need to be
well delocalised, otherwise most of $\langle E_{j}|\mathbf{m}\rangle$
would have almost zero modulus.

Although a coherence time that is as long as $T_{min}$ has been assumed
so far to make the scheme work by letting the signal propagate back
and forth many times, the gateway scheme is also applicable to systems
with short coherence times by modifying it. For example, as shown
in \cite{Maruyama2012}, instead of measuring the spin state in the
accessible area, we may be able to measure in the energy eigenbasis
${|E_{n}\rangle}$, and then the Hamiltonian can be estimated. Such
a global measurement is actually easier in some cases than measuring
the state of a single component. With this modification to the scheme,
however, the graph condition for the accessible area $C$ needs to
be slightly changed; it should be expanded, depending on the graph
structure.

Another potential concern is the (Anderson) localisation. The localisation
of excitation (or spin-up) will take place, if there is too much disorder
in the coupling strengths (see, for example, \cite{Burrell2007}).
Then couplings far away from the controlled region $C$ can no longer
be probed. In turn, this suggests a way of obtaining information on
localisation lengths indirectly. That we cannot `see' beyond the localisation
length would not be a serious problem as our primary purpose is to
identify a quantum system we can control.

When localisation is negligible, the numerical algorithm to obtain
the coupling strengths from the Fourier transform is very stable \cite{Gladwell2004}.
The reason is that the couplings are obtained from a linear system
of equations, so errors in the quantum-state tomography or effects
of noise degrade the estimation only linearly.

Let us also look at the scaling of the problem with the number of
spins. Typically the dispersion relation in one-dimensional systems
of length $N$ is $\cos kN$, which means that the minimal energy
difference scales as $(\Delta E)_{min}\sim N^{-2}$ and thus the total
time interval should be chosen as $T_{min}\sim N^{2}$. This agrees
well with our numerical results tested up to $N=100$. For each sampling
point a quantum-state tomography of a signal of an average height
of $N^{-1}$ needs to be performed. Since the error of tomography
scales inverse proportionally to the square root of the number of
measurements, roughly $N^{2}$ measurements are required for each
tomography.

\section{Quadratic Hamiltonians\label{sec:quadratic}}

So far, we have focused on the Hamiltonians that preserve the total
magnetisation. Nevertheless, it is possible to generalise the above
argument to a more general class. They are those that are quadratic
in terms of annihilation and creation operators, that is
\begin{equation}
H=\sum_{m,n\in E}A_{mn}a_{m}^{\dagger}a_{n}+\frac{1}{2}\left(B_{mn}a_{m}^{\dagger}a_{n}^{\dagger}+B_{mn}^{*}a_{n}a_{m}\right),\label{eq:quadratic_ham}
\end{equation}
which does not preserve the number of quasi-particles $\sum a_{n}^{\dagger}a_{n}$.
Here, $E$ is again the set of interacting nodes as in Eq. (\ref{eq:simple_ham}).
For $H$ to be Hermitian we must have $A=A^{\dagger}$ and $B^{T}=-\epsilon B,$
where $\epsilon=1$ for fermions and $\epsilon=-1$ for bosons, depending
on the particle statistics described by $a$ and $a^{\dagger}$. For
one-dimensional spin chains, the operators $a$ and $a^{\dagger}$
are defined with the standard spin (Pauli) operators through the Jordan-Wigner
transformation \cite{Jordan1928,Lieb1961},
\begin{equation}
a_{n}^{\dagger}a_{n}=\sigma_{n}^{+}\prod_{m<n}Z_{m},\:\:\mathrm{and}\; a_{n}^{\dagger}=\left(\prod_{m<n}Z_{m}\right)\sigma_{n}^{-},\label{eq:jordan_wigner}
\end{equation}
where $\sigma_{n}^{\pm}=(X_{n}\pm iY_{n})/2$. The oparators hereby
defined, $a_{n}$ and $a_{n}^{\dagger},$ satisfy the canonical aniti-commutation
relations for fermions, i.e., $\left\{ a_{m},a_{n}\right\} =0$ and
$\left\{ a_{m},a_{n}^{\dagger}\right\} =\delta_{mn}$. A 1D XX-type
Hamiltonian
\begin{equation}
H=\sum_{m=1}^{N-1}c_{m}[(1+\gamma)X_{m}X_{m+1}+(1-\gamma)Y_{m}Y_{m+1}]+\sum_{m=1}^{N}b_{m}Z_{m}\label{eq:XX-ham}
\end{equation}
with anisotropy factor $\gamma\in[0,1]$ can be rewritten in the form
of Eq. (\ref{eq:quadratic_ham}) through the Jordan-Wigner transformation,
and the matrices $A$ and $B$ will look like
\begin{eqnarray*}
A & = & \left(\begin{array}{cccc}
-2b_{1} & c_{1}\\
c_{1} & -2b_{2} & c_{2}\\
 & c_{2} & -2b_{3}\\
 &  &  & \ddots
\end{array}\right),\:\:\mathrm{and}\; B=\left(\begin{array}{cccc}
0 & \gamma c_{1}\\
-\gamma c_{1} & 0 & \gamma c_{2}\\
 & -\gamma c_{2} & 0\\
 &  &  & \ddots
\end{array}\right).
\end{eqnarray*}
A physically important example of quadratic Hamiltonians is the Ising
chain of spins with transverse magnetic fields, which is expressed
by Eq. (\ref{eq:XX-ham}) with $\gamma=1$ and is relevant for systems,
such as superconducting qubits \cite{Makhlin2001} and NMR. Note also
that once the Hamiltonian of a given system is described in quadratic
form, the operators $a$ and $a^{\dagger}$ can represent not only
fermions, but also bosons by requiring them to obey the bosonic commutation
relations, $[a_{m},a_{n}]=0$ and $[a_{m},a_{n}^{\dagger}]=\delta_{mn}$.
In the following, we shall consider the problem of Hamiltonian tomography
of Eq. (\ref{eq:quadratic_ham}) for 1D chains for simplicity, although
the generalisation to more complex graphs is possible.

Since the Hamiltonian Eq. (\ref{eq:quadratic_ham}) does not preserve
the number of particles, initialising the chain to be $|0...0\rangle$
just by accessing the end node appears to be impossible. Nevertheless,
this difficulty can be circumvented by making use of the propety of
such Hamiltonians. The quadratic Hamiltonian above can be diagonalised
as $H=\sum_{k}E_{k}b_{k}^{\dagger}b_{k}+\mathrm{const.}$ by transforming
$\alpha=(a_{1},...,a_{N},a_{1}^{\dagger},...,a_{N}^{\dagger})^{t}$
into $\beta=(b_{1},...,b_{N},b_{1}^{\dagger},...,b_{N}^{\dagger})^{t}$
as $\beta=T\alpha$, so that operators $b$ and $b^{\dagger}$ still
satisfy the canonical (anti-)commutation relations. So, the quasi-particles
described by $b$ and $b^{\dagger}$ behave as free particles that
almost do not interact with each other.

The `initialisation' works then as follows. Suppose we can initialise
the chain to be in a fixed, but not necessarily known, state $\rho_{0}.$
Though $\rho_{0}$ can be any state, a realistically plausible one
might be a thermal state. We prepare two different states $\psi_{1}$
and $\psi_{2}$ locally at the end site after initialising the chain
to be $\rho_{0}.$ For each initial state we observe the time evolution
at the same end site to get a reduced density matrix $\rho(t|\psi_{i})\,(i=1,2)$
as a function of time. Because the evolution of internal state of
the chain is independent of that of the state at the chain end and
vice versa (thanks to the insensitivity between quasi-particles),
we can extract the pure response of the chain due to the difference
between $\psi_{1}$ and $\psi_{2},$ by comparing $\rho(t|\psi_{1})$
and $\rho(t|\psi_{2})$.

The Hamiltonian of Eq. (\ref{eq:quadratic_ham}) can be rewritten
as
\[
H=\frac{1}{2}\alpha^{\dagger}M\alpha,
\]
where $M$ is a $2N\times2N$ matrix
\begin{equation}
M\equiv\left(\begin{array}{cc}
A & B\\
-\epsilon B^{*} & -\epsilon A^{*}
\end{array}\right),\label{eq:matrix_m}
\end{equation}
with $\epsilon=1$ for fermions and $\epsilon=-1$ for bosons. As
in the previous case of the magnetisation-preserving Hamiltonians,
we assume that all coupling strengths are real and their signs are
known. Also the factor $\gamma=B_{n,n+1}/A_{n,n+1}$ (anisotropy)
is assumed to be constant and known.

Now that we can take it for granted that this $2N\times2N$ matrix
$M$ is symmetric and its entries are all real, a key observation
is to reinterpret $M$ as a Hamiltonian that describes the hopping
of excitations over a graph of $2N$ nodes \cite{Burgarth2011}. That
is, the `Hamiltonian' $M$ preserves the number of excitations in
the $2N$-`spin' network, therefore we can apply the scheme discussed
in previous sections. Of course, the state on which the Hamiltonian
$M$ acts is not a physical spin network, instead it is a fictitious
state represented by a $2N\times1$ vector, $(a_{1},...,a_{1}^{\dagger},...)^{T}.$
So the eigenvectors of $M$ are something different from physical
state vectors.

\begin{figure}
\includegraphics[scale=0.6]{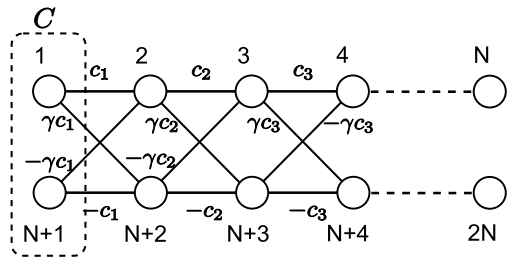}

\caption{\emph{\label{fig:dual_rail_f} }A graph corresponding to the matrix
$M$ with $A$ and $B$ of Eq. (\ref{eq:matrix_m}) and $\epsilon=1$
(fermionic). For bosonic systems, there will be additional edges connecting
nodes $m$ ($1\le m\le N)$ and $N+m$, because $B$ is symmetric,
rather than antisymmetric. For both fermionic and bosonic cases, there
are edges extruding and returning to the same node, corresponding
to the diagonal elements of $A,$ which are not shown here to illustrate
the principal structure of the graph. }
\end{figure}

The graph for a 1D spin chain of Eq. (\ref{eq:XX-ham}) is shown in
Fig. \ref{fig:dual_rail_f}. Accessing the spin 1 in the real chain
corresponds to accessing the nodes 1 and $N+1,$ since what we obtain
from the measurement (and Fourier transform) are the values of $E_{j}$,
$\langle\mathbf{1}|E_{j}\rangle,$ and $\langle\mathbf{N+1}|E_{j}\rangle$\cite{Burgarth2011}.
Here the state $|\mathbf{n}\rangle$ stands for the localised state
on the fictitious $2N$-node graph.

Let us take an Ising chain of $N$ spins with transverse magnetic
fields, i.e., $\gamma=1$ in Eq. (\ref{eq:XX-ham}), as a specific
example to demonstrate how the estimation goes. To make use of the
symmetry the graph in Fig. \ref{fig:dual_rail_f} posesses, let us
define
\[
|n^{\pm}\rangle:=\frac{1}{\sqrt{2}}(|\mathbf{n}\rangle+|\mathbf{N+n}\rangle).
\]
We already have the information about $\langle1^{\pm}|E_{j}\rangle$,
as well as $E_{j}$, from the measurement on the spin 1. The estimation
procedure proceeds as in Sec \ref{sec:The-gateway-scheme}, namely,
by looking at $\langle1^{+}|M|E_{j}\rangle$ we have
\[
E_{j}\langle1^{+}|E_{j}\rangle=-2b_{1}\langle1^{-}|E_{j}\rangle,
\]
whose LHS is known, thus $b_{1}$ can be obtained through the normalisation
condition for $\langle1^{-}|E_{j}\rangle.$ Similarly, evaluating
$\langle1^{-}|M|E_{j}\rangle$ gives
\[
E_{j}\langle1^{-}|E_{j}\rangle=2c_{1}\langle2^{+}|E_{j}\rangle-2b_{1}\langle1^{+}|E_{j}\rangle,
\]
from which $c_{1}$ and $\langle2^{+}|E_{j}\rangle$ can be known.
Also, from $E_{j}\langle2^{+}|E_{j}\rangle=2c_{1}\langle1^{-}|E_{j}\rangle-2b_{2}\langle2^{-}|E_{j}\rangle$
we have $b_{2}$ and $\langle2^{-}|E_{j}\rangle,$ therefore we have
obtained all parameters up to the second spin, so effectively expanded
the accessible area to two spins. Then, this procedure can go on one
by one till we reach the other end of the chain, i.e., the $N$-th
spin, identifying all the paramters in the matrix $M$.

A remark on the initialisation follows. It was shown in \cite{Franco2009}
that, in the case of 1D XX chains of spins-1/2, the estimation of
Hamiltonian parameters is possible without initialising the chain
state. The smart trick there was that the spin 1 was initialised so
that the average value of the $z$-component of spin, i.e., $\langle Z_{1}\rangle$,
was made zero at $t=0.$ The rationale behind it stems from the Jordan-Wigner
transform. Since $a_{n}=\sigma_{n}^{+}\prod_{m<n}Z_{m},$ if we set
$\langle Z_{1}\rangle=0$, the averages of all $a_{n}$ and $a_{n}^{\dagger}\,(n>1)$
at $t=0$ become zero. Their time evolution is expressed as (in the
form of the vector $\alpha$)\foreignlanguage{english}{
\begin{equation}
\alpha_{n}(t)=\sum_{m,k}e^{-iE_{k}t}T_{nk}^{-1}\left(T^{-1}\right)_{km}^{\dagger}\alpha_{m}(0),\label{eq:time_evln_alpha}
\end{equation}
}from which we can see that, in the Jordan-Wigner picture, the initial
state of spins from the second to the $N$-th gives no effect on the
measurement result of the first spin. Here, $T$ is a matrix that
transforms $\alpha$ into $\beta=T\alpha$ as mentioned above to diagonalise
the Hamiltonian. Hence the above initialisation of the first spin
is equivalent to that of the whole chain in the Jordan-Wigner (fermionic)
picture, and thus corresponds to a special case of our description
on initialisation.

\section{Conclusions}

We have seen that despite a severe restriction on our accessibility
a large quantum system can be controlled and its Hamiltonian can be
identified. As a matter of fact, it is unrealistic for any existing
control scheme to have a full access to the system, i.e., a full modulability
for the $d^{2}-1$ parameters for independent Hamiltonians with $d$
being the system dimensionality. In the case of methods based on electron/nuclear
spin resonance, for instance, all we modulate is the external magnetic
field and we do not have a full control over all inter-spin couplings.
Therefore, a guiding theory of quantum control is needed to systematically
understand and design feasible control schemes under a limited access.
The results we have reviewed in this chapter are an example towards
the more generic theory , already showing how powerful a restricted
access can be. Although the limitation for the control in laboratories
would vary, the same or modified methods as what we have seen here
will be of help in making a shortcut towards the realisation of the
full quantum control.

\bibliographystyle{spphys}

\end{document}